\documentclass[prl,twocolumn]{revtex4}  		
\usepackage{graphicx}

\begin{document}

\title
{
Electronic states around a vortex core in high-$T_{c}$ superconductors \\
based on the $t$-$J$ model
}

\author{
Hiroki Tsuchiura$^{1,@}$, Masao Ogata$^{2}$, Yukio Tanaka$^{3}$
and Satoshi Kashiwaya$^{4}$
}
\affiliation{
 CREST, Japan Science and Technology Corporation (JST), 
 Nagoya 464-8603$^{1}$\\
 Department of Physics, University of Tokyo, Bunkyo-ku, 
 Tokyo 113-0033$^{2}$\\
 Department of Applied Physics, Nagoya University, 
 Nagoya 464-8603$^{3}$\\
 NRI of AIST, Umezono, Tsukuba, Ibaraki, 305-8568$^{4}$
} 

\date{\today}

\begin{abstract}
Electronic states around vortex cores in high-$T_{c}$ superconductors
are studied using the two-dimensional $t$-$J$ model in order to treat
the $d$-wave superconductivity with short coherence length and the
antiferromagnetic (AF) instability within the same framework.
We focus on the disappearance of the large zero-energy peak in the local
density of states observed in high-$T_{c}$ superconductors.  
When the system is near the optimum doping, we find that the local
AF correlation develops inside the vortex cores. 
However, the detailed doping dependence calculations confirm that
the experimentally observed reduction of the zero-energy peak is more
reasonably attributed to the smallness of the core size rather than
to the AF correlation developed inside the core.
The correlation between the spatial dependence of the core states
and the core radius is discussed.  
\end{abstract}

\maketitle

There are several experimental results in high-$T_{c}$ superconductors 
which have not been explained in the conventional BCS
$d_{x^{2}-y^{2}}$-wave superconductivity.  
The electronic structure inside the vortex core is one of the most
interesting issues among them.
The conventional theory for $d$-wave vortices based on 
Bogoliubov-de Gennes (BdG) mean-field theory predicts a large and broad peak
at the Fermi energy in the local density of states (LDOS), so-called
zero-energy peak (ZEP), at the vortex core\cite{wang}.
However, scanning tunneling spectroscopy (STS) spectrum in one 
of the high-$T_{c}$ materials, BSCCO,
giving directly the LDOS around the vortex core, shows only a
small-double peak structure at energies $\pm 7$ meV\cite{pan}.
A similar situation was also observed in YBCO compounds\cite{maggio}.

To resolve this discrepancy, there have been proposed several theoretical
attempts, that is, a $d_{x^2-y^2}+s$ state\cite{himeda},
a $d_{x^{2}-y^{2}}+id_{xy}$ state\cite{dxy2,dxy3},
an antiferromagnetic (AF) vortex core\cite{so5,ogata,andersen},
a staggered flux state\cite{kishine},
and a vortex core with small $k_{\rm F}\xi_{0}$\cite{morita,kita} where
$k_{\rm F}$ is the Fermi wave number and $\xi_{0}$ is the coherence length.
These theories are, in greater or lesser degree, based on strongly correlation
effects.
Among them, we consider the AF vortex core and the effects of
small $k_{\rm F}\xi_{0}$ (the small vortex core) because they have
some experimental grounds mentioned below.

The AF vortex core was first predicted using a phenomenological 
Ginzburg-Landau theory\cite{so5} assuming the SO(5) 
symmetry\cite{sczhang}. 
Later this possibility was confirmed in a microscopic 
calculations\cite{ogata,andersen}.
In the $t$-$J$ model, 
it was shown that the AF core is stabilized near the doping rate
$\delta\sim 0.1$, where the system 
is close to the AF instability\cite{himeda2}.
Simultaneously it was observed that the electron density approaches
half-filling inside the AF core\cite{ogata}.
After these calculations, there are several experimental 
studies suggesting the existence of the enhancement of AF
correlations or AF vortex core, such as neutron scattering experiments
in LSCO compounds near optimal doping\cite{lake1,lake2,khay}, 
$\mu$SR\cite{muSR} and NMR\cite{mitro1,kumagai} experiments 
in underdoped or near-optimally doped YBCO.
Therefore it seems natural that the AF vortex core is also realized in
the BSCCO sample\cite{pan} which is optimum doped or slightly
overdoped, 
and that the absence of the large ZEP is due to the mean-field AF gap
(or Mott gap) developed inside the core.
Actually some theoretical studies for the AF vortex core have predicted
the LDOS without the ZEP\cite{ogata,andersen}, or at least, 
large-double peak structure\cite{zhu,berlin}.

The double peak structure is also found in the
small vortex cores (small $k_{F}\xi_{0}$) by using the BdG equations 
for continuum $d$-wave model\cite{morita,kita}. 
In fact, the STM/S data has revealed that a radius for the vortex core
in BSCCO is on the order of 10$\AA$\cite{pan}, which is consistent 
with the short coherence length.
Therefore, the small vortex core is
also a possible explanation for the absence of the ZEP, but
the lattice effects, which play an important role when $k_{F}\xi_{0}$ is small,
need to be clarified.

Motivated by these backgrounds, in this paper we perform a detailed study
on the electronic states around vortex cores in high-$T_{c}$
superconductors, taking account of the discrete lattice effects, 
smallness of the vortex core, and the AF instability.
To investigate their effects within the same framework,
we use the $t$-$J$ model
which has a phase diagram consistent with experiments\cite{yoko-ogata}
and enables us to study the reasonable doping dependence.
We find that the reduction of ZEP is more
reasonably attributed to the smallness of the core size rather than
to the AF correlation developed inside the core.
It is remarkable that a simple-minded picture of AF core does not 
play important roles in the suppression of LDOS.

The Hamiltonian of the $t$-$J$ model is written as
\begin{eqnarray}
{\mathcal H} &=& -\sum_{\langle i,j \rangle,\sigma}
P_{G}( t_{ij}c^{\dag}_{i\sigma} c_{j\sigma} + {\rm h.c.} )P_{G}
 \nonumber \\
& & + J\sum_{\langle i,j\rangle}\mbox{\boldmath $S_{i}\cdot S_{j}$}
- \mu\sum_{i,\sigma}c^{\dag}_{i\sigma} c_{i\sigma}
\label{hamil}
\end{eqnarray}
in the standard notation where $\langle i,j\rangle$ means the summation
over nearest-neighbor pairs.
The Gutzwiller's projection operator $P_{G}$ is defined as
$P_{G} = \Pi_{i}( 1 - n_{i\uparrow}n_{i\downarrow})$.
The uniform magnetic field is introduced in terms of Peierls phase
of the hopping term as
$t_{ij} = t\exp\left( \frac{ie}{\hbar c}
\int_{i}^{j}\mbox{\boldmath $A\cdot dr$}\right)$\cite{wang,himeda}.
The BdG equation based on the extended Gutzwiller
approximation is 
\begin{equation}
\left(
\begin{array}{cc}
H_{ij}^{\uparrow} & F_{ij} \\
F_{ji}^{*} & -H_{ji}^{\downarrow}
\end{array}
\right)
\left(
\begin{array}{c}
u_{j}^{\alpha} \\
v_{j}^{\alpha}
\end{array}
\right)
= E^{\alpha}
\left(
\begin{array}{c}
u_{i}^{\alpha} \\
v_{i}^{\alpha}
\end{array}
\right)
,
\end{equation}
with
\begin{eqnarray}
H_{ij}^{\sigma} 
&=& -\sum_{\tau}\left( t_{ij}^{\rm eff} + J_{ij}^{\rm eff}\tilde{\chi}_{ji}
\right) \delta_{j,i+\tau}  \nonumber \\
& & + \sigma\delta_{ij}\sum_{\tau}h_{i,i+\tau}^{\rm eff}
 - \mu\delta_{ij} ,
\nonumber \\
 F_{ij}^{*} &=& -\sum_{\tau} J_{ij}^{\rm eff}\tilde{\Delta}_{ij}
 \delta_{j,i+\tau}  ,
\end{eqnarray}
where $i+\tau$ represents the nearest neighbor sites of the 
site $i$, and $\sigma = \pm 1$.
The renormalized parameters $t_{ij}^{\rm eff}$, $J_{ij}^{\rm eff}$ and 
$h_{ij}^{\rm eff}$ are determined from cluster calculations\cite{ega} 
which reproduce the variational Monte Carlo results\cite{himeda2}.
They have the following forms:
\begin{eqnarray}
t_{ij}^{\rm eff} &=& g_{t}(i,j)t_{ij}, ~~
J_{ij}^{\rm eff} = \frac{1}{2}g_{s}^{xy}(i,j)J
 + \frac{1}{4}g_{s}^{z}(i,j)J , 
\nonumber \\
h_{ij}^{\rm eff} &=& \frac{1}{2}g_{s}^{z}(i,j)Jm_{j}
 + \frac{\partial\langle H_{ij}\rangle}{\partial m_{i}} ,
\label{renorm}
\end{eqnarray}
where $g_t(i,j), g_s^{xy}(i,j)$ and $g_s^z(i,j)$ depend on the 
local expectation values 
$\Delta_{ij}=\frac{1}{2} \langle c_{i\uparrow}^{\dag}c_{j\downarrow}^{\dag}
 - c_{i\downarrow}^{\dag}c_{j\uparrow}^{\dag}\rangle$, 
$\chi_{ij} = \langle c_{i\sigma}^{\dag}c_{j\sigma}\rangle$, and
$m_i=\frac{1}{2}\langle n_{i\uparrow}-n_{i\downarrow}\rangle$\cite{foot}.
For example, 
\begin{equation}
g_s^{xy}(i,j) = \left( \frac{2(1-\bar{\delta}_{ij})}
 {1-\bar{\delta}_{ij}^2+4\bar{m}_{ij}^2}\right)^2
 \cdot a_{ij}^{-7} ,
\end{equation}
where $\bar{\delta}_{ij}$ ($\bar{m}_{ij}$) is an average of hole 
density (AF moment) for the corresponding bond and $a_{ij}$ is a 
factor close to one which depends on $\Delta_{ij}, \chi_{ij}, 
\bar{\delta}_{ij}$ and $\bar{m}_{ij}$\cite{ega}.
These somewhat complicated renormalized parameters are necessary 
for treating the AF and d-wave superconductivity in the same 
framework.  
If one uses simple slave-boson mean-field theories\cite{inaba},
the AF order parameter is overestimated and thus one cannot 
discuss the AF vortex core because the AF order exists even in the bulk.  
In the present method, AF order does not exist away from the core 
for $\delta\ge 0.10$, which is consistent with the numerical 
results\cite{himeda2}.

To make the Peierls phase compatible with vortex lattice symmetry,
we need a magnetic unit cell with $2N\times N$ sites including 
two vortices.
In this case, by the appropriate choice of gauge\cite{wang,himeda}, 
the order parameter $\Delta_{ij} $ has a translational symmetry
with respect to the magnetic unit cell. 
Here we take $N=$22-26.
We solve numerically the BdG equation and carry out an iteration until
$\Delta_{ij}, \tilde{\Delta}_{ij}, \chi_{ij}, \tilde{\chi}_{ij}, 
n_{i\sigma}$ are determined self-consistently.
We take $J/t = 0.3$ and examine the various values of the doping rates
$\delta = 0.10$-$0.15$, that is, from near optimum doped to overdoped
region.

\begin{figure}[htb]
\begin{center}
\includegraphics[clip,width=4.5cm]{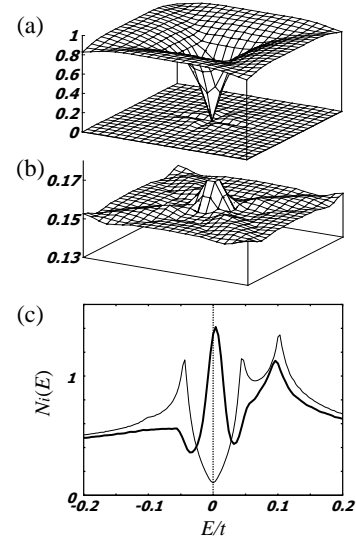}
\end{center}
\vskip -4mm
\caption{
(a) Spatial dependence of the amplitude of $d$-wave (upper) and $s$-wave
(lower) components of the superconducting order parameter around
the vortex core. The doping rate is $\delta = 0.15$.
(b) The hole density $n_{i}^{h}$.
(c) The LDOS obtained at the vortex core center (the thick line) 
and obtained without the magnetic field (the thin line).
}
\label{fig1}
\end{figure}
First, let us look at the results for the overdoped region.
Figure \ref{fig1} shows the superconducting order parameters,
the hole density, and the LDOS obtained around the vortex core for
$\delta = 0.15$.
We can see the common features found in several theoretical studies
within the weak-coupling $d$-wave superconductivity\cite{wang}, 
that is, fairly large vortex cores, slightly induced $s$-wave 
component\cite{himeda}, the increase of the hole density inside 
the core, and the ZEP in the LDOS.
We note that the local AF order inside the core does not exist in this case.

\begin{figure}[htb]
\begin{center}
\includegraphics[clip,width=4.5cm]{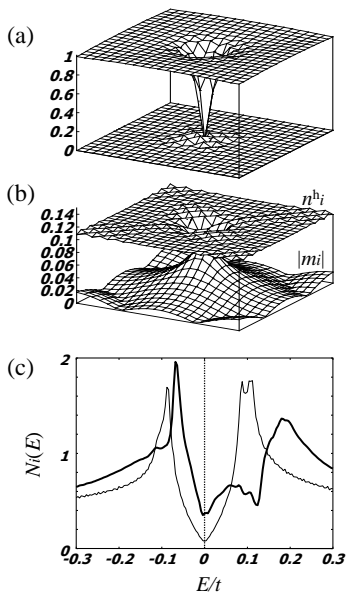}
\end{center}
\vskip -4mm
\caption{
(a) Spatial dependence of the amplitude of $d$-wave (upper) and $s$-wave
(lower) components of the superconducting order parameter around
the vortex core. The doping rate is $\delta = 0.11$.
(b) The hole density $n_{i}^{h}$ (upper) and local antiferromagnetic
moment (lower).
(c) The LDOS obtained at the vortex core center (the thick line) 
and obtained without the magnetic field (the thin line).
}
\label{fig2}
\end{figure}
The situation drastically changes when the system approaches the
optimum doping.
Figure 2 shows the results for $\delta = 0.11$ in the same manner as
in Fig.\ \ref{fig1}.
Similar results for $\delta = 0.10$ have been shown before\cite{ogata},
but here much larger system-size is achieved.  From Fig.\ \ref{fig2}, 
we can see that the size of the vortex
core is fairly small, that is, the core having a radius of about 3
lattice spacings is realized.
The magnitude of the induced $s$-wave component is
also small as in the case of $\delta = 0.15$.
Figure \ref{fig2}(b) shows the spatial dependences of the local AF moment
$|m_{i}|$ and of the hole density $n^{h}_{i}$.
It is found that AF correlation develops and the hole density decreases
inside the vortex core, in sharp contrast with the
$\delta = 0.15$ case shown in Fig.\ \ref{fig1}.
We should note here that the AF moment extends outside the core, 
which is consistent with experiments. 

The LDOS obtained at the center of the vortex core is shown in
Fig.\ \ref{fig2}(c).
We can see that the LDOS shows neither the ZEP nor an explicit
double-peak structure.
%
\begin{figure}[htb]
\begin{center}
\includegraphics[clip,width=4.5cm]{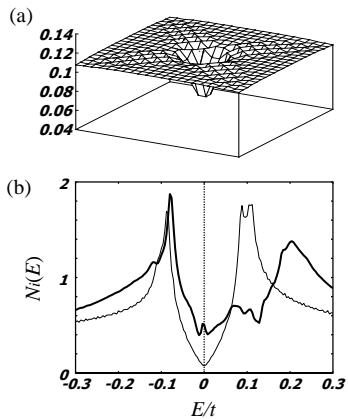}
\end{center}
\vskip -4mm
\caption{
The hole density $n_{i}^{h}$ (a) and the LDOS (b) around the vortex core
obtained by solving the BdG equations {\it without} AF order parameter.
The doping rate is $\delta = 0.11$.
}
\label{fig3}
\end{figure}
To clarify the effect of the local AF moment on the LDOS, 
we solve the BdG equations {\it without} AF order parameter for the
same doping rate $\delta = 0.11$, as shown in Fig.\ \ref{fig3}.
The LDOS obtained here (Fig.\ \ref{fig3}(b)) is quite similar to that
shown in Fig.\ \ref{fig2}(c), but at this time, a small peak can be
found at $E\sim 0$.
This peak is regarded as a trace of the ZEP. 
The only effect of the presence of AF order parameter is to 
annihilate the small peak at $E\sim 0$. 
We have confirmed that the spatial dependence of the amplitude of $d$-
and $s$-wave components, and the hole density (Fig.\ \ref{fig3}(a)) are
quite similar to those shown in Fig.\ \ref{fig2}(a) and (b).  

From this comparison, we speculate as follows.
The overall reduction of the spectral weight inside the vortex core 
already occurs without AF order parameters.  
Therefore this reduction is not due to the presence of the AF moment 
but due to the smallness of the core.
For this doping rate ($\delta = 0.11$), the hole density inside the
core decreases.
This causes the appearance of the AF vortex core found in
Fig.\ \ref{fig2}(b), but the effect of the AF core is subsidiary 
for the LDOS.  

\begin{figure}[htb]
\begin{center}
\includegraphics[clip,width=4.5cm]{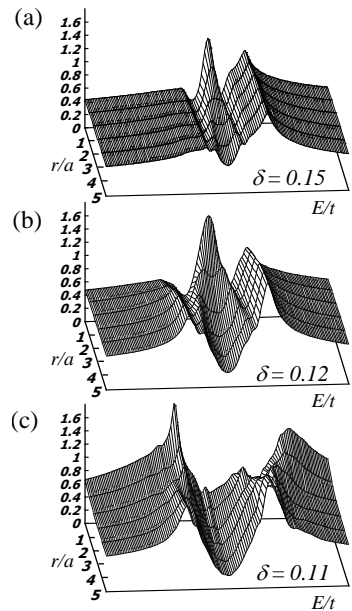}
\end{center}
\vskip -4mm
\caption{
The LDOS as a function of energy $E/t$ and distance $r/a$ from the
vortex core center, where $a$ is the lattice constant.
The range of the energy is $-0.3\leq E/t\leq0.3$.
}
\label{fig4}
\end{figure}

In order to clarify the origin of the reduced spectral
weight of ZEP due to the smallness of the core, 
we study the spatial dependence of the
LDOS for several hole dopings.
Figure \ref{fig4} shows the LDOS at $\delta = 0.15, 0.12$ and $0.11$
as a function of energy $E/t$ and distance $r/a$ from the vortex core
center, with $a$ being the lattice constant.
Here the distance dependence is plotted from the vortex core center
($r=0$) along the (1,0) direction of the square lattice.
First we look at the result for $\delta = 0.15$ shown in Fig.\ \ref{fig4}(a).
As $r/a$ increases from 0, the ZEP first becomes broader and smaller,
then split into two peaks for $r/a \geq 3$.
This kind of behavior of the LDOS, that is, the splitting of the ZEP
away from the vortex core center, has been reported for $s$-wave vortex
cores by Gygi and Schl\"{u}ter\cite{gygi}.
They have shown that, in the $s$-wave case, the two peaks found 
away from the vortex core center correspond to the electronic bound states 
(core states) having
higher {\it angular momentum} $\mu$.
The angular momentum is quantized such that $\mu = m+\frac{1}{2}$ where
$m$ is an integer.
For small $\mu$, the energy eigenvalues of the core states are known
to have the approximate dispersion relation $E_{\mu}\sim\omega_{0}\mu$
where $\omega_{0}$ is an energy quantum\cite{caroli}.
In a semiclassical sense, the vortex core states have maximum amplitude 
at a distance $r\sim \mu/p_{\rm F}$ from the vortex core center.
Of course in $d$-wave superconductors, the angular momentum is not
conserved in a strict sense because axial symmetry is broken.
Quite recently, however, it has been pointed out that there is an alternative
quantum number in $d$-wave superconductors which corresponds to the angular 
momentum\cite{koyama}.
Therefore the distance dependence of the splitting amplitude of ZEP
shown in Fig.\ \ref{fig4} indicates the dispersion relation of the core states:
i.e., $E_{\mu}\sim\omega_{0}\mu$.
At $\delta = 0.12$ shown in Fig.\ \ref{fig4}(b), the splitting of the
ZEP exists even at $r/a = 1$, that is, the ZEP can be seen only
at the vortex core center.
Note that the AF moment is not induced for this case.
Comparing the results in Fig.\ \ref{fig4}(a) and (b), we find that
the slope of the dispersion relation becomes greater as the doping rate
$\delta$ decreases.  
This greater slope corresponds to larger
$\omega_{0}$ or larger interlevel spacing of the core states, which is 
probably caused by the decrease of the core size. 
As the doping rate decreases further ($\delta = 0.11$), the LDOS around
$E\sim 0$ is significantly reduced as shown in 
Fig.\ \ref{fig4}(c)\cite{foot2}.

Finally let us comment on the effects of the next-nearest-neighbor hopping
$t'$.
The critical value of $\delta$ where the AF core appears depends on
the value of $t'$ which reinforces our viewpoint. 
For some values of $t'$, the AF core is realized even when the size of
the vortex core is not so small.
In this case, the spectral weight of the ZEP is not reduced owing to
the large size of the vortex core, and there appears a large double peak
structure in the LDOS caused by the presence of the AF 
moment\cite{zhu,berlin}.
This, however, does not explain the reduction of the spectral weight 
inside the core observed in YBCO or BSCCO.

In summary, we have investigated the electronic states around vortex
cores in high-$T_{c}$ superconductors based on the $t$-$J$ model.
The induced local AF moment and the LDOS without ZEP are found inside
the vortex cores.
However, the origin of the absence of ZEP is mainly due to the smallness
of the vortex core sizes, which can be understood within the conventional
BdG theory with short coherence length.  
We find that the doping dependence is very important and thus it is
quite interesting to study the overdoped region or low-$T_c$
superconductors experimentally, in which the vortex core size is large
and the ZEP with the spatial dependence as shown in Fig.\ 4(a) 
will be observed.

The authors wish to thank J.\ Inoue and H.\ Ito for their useful discussions.

%



\begin{thebibliography}{99}
\bibitem{foot2}
We notice that the sudden change of the LDOS is due 
to the appearance of staggered flux state inside the core 
which is caused by the smallness of the core and reduction of the 
hole density.  This is consistent with the recent work 
by P.\ A.\ Lee (private communication).
\end{thebibliography}
\end{document}